\documentclass[%
reprint,
superscriptaddress,
preprintnumbers,
nofootinbib,
amsmath,amssymb,
aps,
pra,
]{revtex4-1}

\usepackage{graphicx}
\usepackage{dcolumn}
\usepackage{grffile}
\usepackage{dcolumn}
\usepackage{bm}
\usepackage{epsfig}
\usepackage{mathrsfs}  
\usepackage{subfigure}
\usepackage{multirow}
\usepackage{epstopdf}
\usepackage{amsmath}
\usepackage{algorithmicx}
\usepackage{amssymb}
\usepackage{tensor}
\usepackage[math]{cellspace}
\usepackage{bookmark}
\usepackage[usenames,dvipsnames]{xcolor}
\usepackage{hyperref}
\usepackage{slashed}
\usepackage{youngtab}
\usepackage{physics}
\usepackage{tensor}
\usepackage[capitalise]{cleveref}
\usepackage{orcidlink}

\crefformat{equation}{Eq.~(#2#1#3)}
\crefformat{table}{Tab.~(#2#1#3)}
\crefformat{figure}{Fig.~(#2#1#3)}
\crefformat{appendix}{App.~(#2#1#3)}
\crefformat{section}{Sec.~(#2#1#3)}
\crefformat{subsection}{Subsec.~(#2#1#3)}

\makeatletter
\renewcommand*\env@matrix[1][\arraystretch]{%
  \edef\arraystretch{#1}%
  \hskip -\arraycolsep
  \let\@ifnextchar\new@ifnextchar
  \array{*\c@MaxMatrixCols c}}
\makeatother

\allowdisplaybreaks
\begin{document}

\preprint{YITP-26-41, RESCEU-12/26, IPMU26-0017}

\title{
Quantum mechanics with a ghost:
Counterexamples to spectral denseness
}

\author{C\'edric~Deffayet\,\orcidlink{0000-0002-1907-5606}}
\email{cedric.deffayet@ens.fr}
\affiliation{
Laboratoire de Physique de l'\'Ecole normale sup\'erieure, ENS, Universit\'e PSL, CNRS, Sorbonne Universit\'e, Universit\'e Paris Cit\'e, F-75005 Paris, France
}

\author{Atabak Fathe Jalali\,\orcidlink{0009-0007-7717-3101}}
\email{jalali@fzu.cz}
\affiliation{
CEICO--Central European Institute for Cosmology and Fundamental Physics, FZU--Institute of Physics of the Czech Academy of Sciences, Na Slovance 1999/2, 182 00 Prague 8, Czech Republic
}
\affiliation{
Institute of Theoretical Physics, Faculty of Mathematics and Physics, Charles University, V Hole\v{s}ovi\v{c}k\'ach 2, 180 00 Prague 8, Czech Republic
}

\author{Aaron~Held\,\orcidlink{0000-0003-2701-9361}}
\email{aaron.held@phys.ens.fr}
\affiliation{
Institut de Physique Théorique Philippe Meyer, Laboratoire de Physique de l’\'Ecole normale sup\'erieure (ENS), Universit\'e PSL, CNRS, Sorbonne Universit\'e, Universit\'e Paris Cité, F-75005 Paris, France
}

\author{Shinji Mukohyama\,\orcidlink{0000-0002-9934-2785}} 
\email{shinji.mukohyama@yukawa.kyoto-u.ac.jp}
\affiliation{
Center for Gravitational Physics and Quantum Information, Yukawa Institute for Theoretical Physics, Kyoto University, 606-8502, Kyoto, Japan
}
\affiliation{
Research Center for the Early Universe (RESCEU), Graduate School of Science, The University of Tokyo, Hongo 7-3-1, Bunkyo-ku, Tokyo 113-0033, Japan
}
\affiliation{
Kavli Institute for the Physics and Mathematics of the Universe (WPI), The University of Tokyo Institutes for Advanced Study, The University of Tokyo, Kashiwa, Chiba 277-8583, Japan
}

\author{Alexander Vikman\,\orcidlink{ 0000-0003-3957-2068}} 
\email{vikman@fzu.cz}
\affiliation{
CEICO--Central European Institute for Cosmology and Fundamental Physics, FZU--Institute of Physics of the Czech Academy of Sciences, Na Slovance 1999/2, 182 00 Prague 8, Czech Republic
}

\begin{abstract}
We quantise integrable point-particle systems with opposite-sign kinetic terms and nontrivial interactions. Using methods from separability theory, we show that previously determined classical stability conditions also imply discrete separated eigenvalue spectra. The resulting energy spectrum is unbounded above and below but not necessarily dense. We establish sufficient conditions for (i) exactly one accumulation point, or (ii) none at all. This dispels the widespread notion that ghostly quantum systems must have a continuous or dense energy spectrum.
\end{abstract}

\maketitle

\noindent{\textbf{Introduction.}}
Dynamical degrees of freedom with negative kinetic energy -- commonly referred to as ghosts -- are widely dismissed as unphysical, see~e.g.~\cite{Woodard:2015zca}. 
Vice versa, the absence of ghosts is often elevated to a ``folk no-go theorem'' and serves as a key construction principle throughout all areas of fundamental physics~\cite{Copeland:2006wr,Sotiriou:2008rp,Clifton:2011jh,deRham:2014zqa,Berti:2015itd}.
This dismissal of ghosts is rooted in the unboundedness of the total Hamiltonian in phase space, which no longer imposes a \emph{kinematical} bound. In turn, this has led to the expectation that opposite-sign excitations will inevitably lead to a \emph{dynamical} instability. While it is straightforward to construct examples which realise this expectation, there is no proof that the above inference from kinematics to dynamics is inevitable.

To the contrary, recent work on counterexamples~\cite{Deffayet:2021nnt,Deffayet:2023wdg} suggests that an effective decoupling outside of a finite region in configuration space provides a generic dynamical mechanism to evade the ``folk no-go theorem'' -- at least in the context of classical time evolution. 
Here, we will extend these counterexamples to the quantum world.
In particular, it has been established that a class of integrable point-particle systems with opposite-sign kinetic terms exhibits bounded motion for all initial conditions~\cite{Deffayet:2021nnt,Deffayet:2023wdg}. 
While the proof relies on integrability, numerics suggest that the same dynamical decoupling mechanism holds also in absence of integrability~\cite{Deffayet:2023wdg}. Much is also starting to be understood about classical field theory~\cite{Deffayet:2025lnj,Held:2025fii,Figueras:2025gal}: We comment on this in the discussion.

Here, we remain with a finite number of degrees of freedom and focus on the ghostly Hamiltonian
\begin{align}
    \mathcal{H} &= 
    \frac{p_x^2}{2} - \frac{p_y^2}{2}
    + V(x,y)
    \;,
    \label{eq:ghostly-Hamiltonian}
\end{align}
where $(x,y,p_x,p_y)$ denote position and momenta. We canonically quantise by the standard prescription
\begin{align}
    p_x \to \hat{p}_x = -i\hbar\partial_x\;,
    \quad
    p_y \to \hat{p}_y = -i\hbar\partial_y\;,
    \label{eq:standard-quantisation}
\end{align}
such that the stationary Schr\"odinger equation reads
\begin{align}
    \hat{\mathcal{H}}\,\Psi
    =
    \left[
        -\frac{\hbar^2}{2}\left(\partial_x^2 - \partial_y^2\right)
        +V(x,y)
    \right] \Psi(x,y)
    =
    E\,\Psi
    \;.
\label{eq:Sch_xy_start}
\end{align} 
The quantum theory is defined on the standard Hilbert space
$\mathscr H_{xy}\,{=}\,L^2(\mathbb R^2,dx\,dy)$ and probabilities are computed with respect to the measure $dx\,dy$.
The purpose of this letter is to dispel the common expectation that the energy spectrum of any such ghostly system is continuous or dense (see, e.g. \cite{Smilga:2008pr,Smilga:2017arl,Damour:2021fva} for decoupled oscillators).
To do so, we restrict to potentials $V(x,y)$ which exhibit two separate $\mathbb{Z}_2$ reflection symmetries $x\,{\mapsto\,}{-}x$ and $y\,{\mapsto\,}{-}y$ and, moreover, an additional constant of motion (see~\cite{Deffayet:2023wdg} and below). 
This allows us to construct examples with a discrete energy spectrum.
\\[-0.7em]

\noindent{\textbf{Integrability and separable coordinates.}}
We focus on a class of integrable systems commonly denoted in separable coordinates $(\alpha,\beta)$. The original system with equal-sign kinetic terms is due to Liouville~\cite{Liouville1846} and was extended to the opposite-sign case in~\cite{Deffayet:2023wdg}.
The coordinate transformation is most conveniently denoted in two steps $(x,y)\,{\mapsto}\,(u,v)\,{\mapsto}\,(\alpha,\beta)$. The intermediate coordinates\footnote{Our definition of $v^2$ differs by a minus sign in comparison to~\cite{Deffayet:2023wdg}.} are confocal (elliptic--hyperbolic) coordinates $(u,v)$ with focal parameter $c<0$ which we set to $c=-1$, i.e.,
\begin{align}
	x^2 = u^2v^2\;,
\qquad
	y^2 = (u^2 + 1)(v^2 - 1)\;,
\label{eq:xytouv}
\end{align}
up to quadrant sign choices in $(x,y)$. The $\mathbb{Z}_2$ symmetries of the model uniquely extend all definitions from one quadrant to the full $(x,y)$-plane. 
The separable coordinates are then obtained as
\begin{align}
	u=\sinh\alpha\;,
	\qquad 
	v=\cosh\beta\;.
\end{align}
The physical coordinate domain corresponds to $\alpha{\in}(0,\infty)$ and $\beta{\in}(0,\infty)$ (again, up to quadrant choices).

In $(\alpha,\beta)$ coordinates, the classical Hamiltonian $\mathcal{H}$ and the additional conserved quantity $\mathcal{I}$ take the form
\begin{align}
    \mathcal{H}
    &=
    \frac{p_\alpha^2/2-p_\beta^2/2}{u(\alpha)^2+v(\beta)^2}
    +\frac{f(u(\alpha))-g(v(\beta))}{u(\alpha)^2+v(\beta)^2},
    \label{eq:HLV_classical_alphabeta}
    \\
    \mathcal{I}
    &=
    p_\alpha^2+2\,f(u(\alpha))-2\,u(\alpha)^2\,\mathcal{H},
    \label{eq:ILV_classical_alphabeta}
\end{align}
with two free functions $f,g$.
We note that the respective configuration space metric $ds^2\,{=}\,(u^2+v^2)(d\alpha^2 - d\beta^2)$ is flat and of definite sign, i.e., $u^2+v^2\,{\geqslant}\,1$ throughout the domain, including in particular the inner boundaries, e.g., $u(\alpha)^2+v(\beta)^2\stackrel{\alpha,\beta\to 0}{\longrightarrow} 1$.

As shown in~\cite{Deffayet:2023wdg}, this integrable class contains a subclass, for which the potential $V$ is polynomial in $(x,y)$,
\begin{align}
    f(u)=\sum_{i=1}^N\,\mathcal{C}_i\,u^{2i}\;,
    \quad
    g(v)=\sum_{i=1}^N\,(-1)^i\mathcal{C}_i\,v^{2i}\;,
    \label{eq:polynomial-subclass-fg}
\end{align}
with parameters $\mathcal{C}\,{=}\,(\mathcal{C}_1,\mathcal{C}_2,\dots,\mathcal{C}_N)$. This polynomial subclass will serve as an instructive example.
\\[-0.7em]

\noindent{\textbf{Classical separability.}}
In coordinates $(\alpha,\beta)$, the classical motion is separable.
This follows from general theorems~\cite{stackel1893ueber} but can also be made explicit as follows.
We can solve $\mathcal{H}=E$ (energy) and $\mathcal{I}=I$ (hidden constant of motion) with respect to $p_\alpha^2$ and $p_\beta^2$ and find
\begin{align}    
    p_\alpha^2/2 &= - \big(
        f(u(\alpha)) - E\,u(\alpha)^2 - I/2
    \big)\;,
    \\*
    p_\beta^2/2 &= - \big(
        g(v(\beta)) + E\,v(\beta)^2 - I/2
    \big)\;.
\end{align}
Physical motion is restricted to a region in which the momenta are real-valued, hence, the right-hand sides are positive. Conversely, the motion is confined to finite configuration space variables $(\alpha,\beta)$ if the right-hand sides turn negative at sufficiently large $|u|$ and $|v|$, respectively. Global stability is thus guaranteed if the former holds for any choice of $E$ and $I$. It is thus sufficient if the functions are continuous and $f(u) - Eu^2\,{\to}\,\infty$ for large $|u|\,{\to}\,\infty$ as well as $g(v) + Ev^2\,{\to}\,\infty$ for large $|v|\,{\to}\,\infty$. These conditions are in agreement with those obtained in~\cite{Deffayet:2023wdg} and considerably simplify a proof of global stability. 
We will now see that the same conditions re-appear also in the quantised system.
\\[-0.7em]

\noindent{\textbf{Quantum separability.}}
Due to Robertson~\cite{robertson1928bemerkung}, Eisenhart~\cite{eisenhart1934separable} and Benenti--Chanu--Rastelli~\cite{benenti2002remarks}, it is known that, for St\"ackel systems~\cite{stackel1893ueber} with a diagonal Ricci tensor on configuration space (such as the one of interest here), classical separability implies quantum separability.
Once again, this is part of general separability theory but can also be made explicit as follows. 
Under the transformation $(x,y)\mapsto(\alpha,\beta)$, the configuration-space metric is conformally flat and the kinetic operator transforms to
\begin{align}
    \partial_x^2-\partial_y^2
    =
    \frac{1}{u(\alpha)^2 + v(\beta)^2}\,
    \left(\partial_\alpha^2 - \partial_\beta^2\right)
    \;,
\label{eq:box_transform}
\end{align}
i.e., no first-derivative terms appear.\footnote{Equivalently, one may quantise in curvilinear coordinates using the Laplace--Beltrami prescription~\cite{podolsky1928quantum,dewitt1952point,carter1977killing,essen1978quantization,blaszak2019quantum}. For the conformally flat metric in 2D, this yields the same separated equations.}
Further, the potential takes St\"ackel form (see~\cref{eq:HLV_classical_alphabeta}), i.e., exhibits the same denominator.
Substituting~\cref{eq:box_transform} and the potential in~\cref{eq:HLV_classical_alphabeta} into \cref{eq:Sch_xy_start}, we obtain
\begin{align}
    \left[
        - \frac{\hbar^2}{2}\,\frac{\partial_\alpha^2-\partial_\beta^2}{u^2 + v^2}
        + \frac{f-g}{u^2 + v^2}
    \right]
    \Psi(\alpha,\beta)
    =
    E\,\Psi(\alpha,\beta)
    \;.
    \label{eq:Schroedinger_alphabeta}
\end{align}
Given the Jacobian $dx\,dy\,{=}\,(u^2+v^2)\,d\alpha\,d\beta$, \cref{eq:Schroedinger_alphabeta} is Hermitian with respect to the measure
$(u^2+v^2)\,d\alpha\,d\beta$.
Multiplying with $(u^2 + v^2)$ gives a manifestly separable form and a product ansatz $\Psi(\alpha,\beta)\,{=}\,\phi(\alpha)\chi(\beta)$ with a separation constant $\lambda$ separates the equation into
\begin{align}
    - \frac{\hbar^2}{2}\phi''(\alpha) 
    + \underbrace{\left[f(u(\alpha)) - E\,u(\alpha)^2\right]}_{\equiv V_u(u(\alpha))}\,\phi(\alpha)
    &=
    \lambda\phi(\alpha)
    \;,
    \notag\\[-0.7em]    
    - \frac{\hbar^2}{2}\chi''(\beta) 
    + \underbrace{\left[g(v(\beta)) + E\,v(\beta)^2\right]}_{\equiv V_v(v(\beta))}\,\chi(\beta)
    &=
    \lambda\chi(\beta)
    \;.
    \label{eq:separated}
\end{align}
The decoupled effective Schr\"odinger potentials $V_u(u(\alpha))$ and $V_v(v(\beta))$ are bounded from below, diverge at the outer boundaries, i.e., $V_u\!\stackrel{\alpha\to\infty}{\longrightarrow}\!\infty$ and $V_v\!\stackrel{\beta\to\infty}{\longrightarrow}\!\infty$ and are finite elsewhere, whenever the classical stability conditions (see above) are fulfilled.

The inner boundaries $\alpha\,{\to}\,0^+$ and $\beta\,{\to}\,0^+$ correspond to the coordinate axes $x\,{=}\,0$ and $y\,{=}\,0$, respectively.
Since $V(x,y)$ is even in $x$ and $y$, the full-plane Hilbert space
$\mathscr{H}_{xy}\,{=}\,L^2(\mathbb R^2,dx\,dy)$
decomposes into four orthogonal parity sectors $\mathscr{H}_{xy}=\mathscr{H}^{++}\oplus \mathscr{H}^{+-}\oplus \mathscr{H}^{-+}\oplus \mathscr{H}^{--}$, each separately invariant under time evolution.
Restricting a parity-definite wave function to the first quadrant yields a unitary identification with the quadrant Hilbert space
$L^2(\mathbb R_+^2,dx\,dy)$, where the parity fixes the boundary condition on the corresponding axis:
Even/odd parity corresponds to Neumann/Dirichlet boundary conditions.

Each separated~\cref{eq:separated} then defines a self-adjoint and confining one-dimensional Schr\"odinger operator, hence, has a discrete spectrum. We denote the $m^{th}$ (respectively $n^{th}$) eigenvalue by $\lambda_m^{(u)}(E)$ (and $\lambda_n^{(v)}(E)$).
Energy quantisation follows from
demanding that
\begin{align}
    \Delta_{mn}(E) \equiv \lambda_m^{(u)}(E) - \lambda_n^{(v)}(E)
    \label{eq:energy-quantisation-condition}
\end{align}
vanishes. In fact, this is the only remaining place, where the two decoupled equations ``talk to each other''. 
\\[-0.7em]

\begin{figure}
    \includegraphics[width=\linewidth]{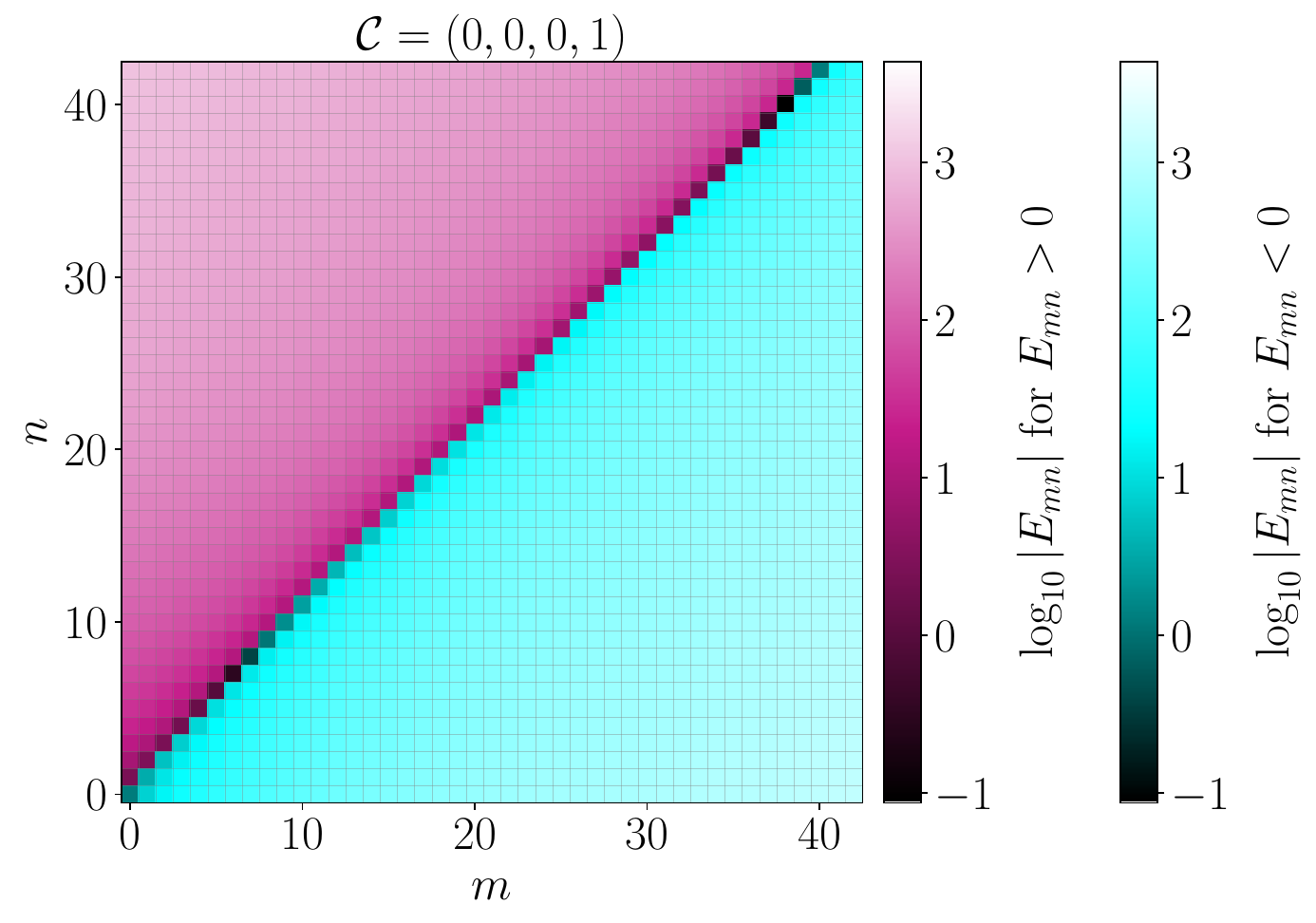}
    \\[-1.2em]
    \caption{
    A generic example ($\mathcal{C}{=}(0,0,0,1)$) of the discrete, equal-parity (even) energy levels $|E_{mn}|$ (with colour indicating the sign) in the $(m,n)$ plane. For large $m$ or $n$ away from the diagonal, the energy-eigenvalues become increasingly sparse. Repeated near-zero accumulation can only occur along sequences for which $(m_k{-}n_k){=}\text{const}$, asymptotically.
    }
    \label{fig:generic-example}
\end{figure}
%

\noindent{\textbf{Energy spectrum.}}
The energy spectrum is not bounded, cf.~\cref{fig:generic-example}, neither above nor below, but for each $m$ and $n$ there exists a unique $E_{mn}$.
Formally, a proof of this relies on the Hellmann--Feynman theorem~\cite{hellman1937einfuhrung,feynman1939forces}, which relates the parameter dependence of a self-adjoint operator to its eigenvalues via expectation values of its eigenstates.
Here, the theorem, combined with $\langle
v^2\rangle\,{=}\,\langle\cosh^2\beta\rangle\,{\geqslant}\,1$, implies that $\partial_E\Delta_{mn}(E)\,{=}\,{-}\,(\langle u^2\rangle\,{+}\,\langle v^2\rangle)\,{\leqslant}\,-1$. It follows that $\Delta_{mn}(E)$ is strictly decreasing and that $\Delta_{mn}(E)\,{\to}\,{\mp}\,\infty$ as $E\,{\to}\,{\pm}\,\infty$.
By continuity, $\Delta_{mn}(E)\,{=}\,0$ occurs exactly once. We denote the unique root of $\Delta_{mn}(E)$ as $E_{mn}$ and correspondingly
$\lambda_m^{(u)}(E_{mn}){=}\lambda_n^{(v)}(E_{mn})$ as $\lambda_{mn}$.

\begin{figure}
    \centering
    \includegraphics[width=0.97\linewidth]{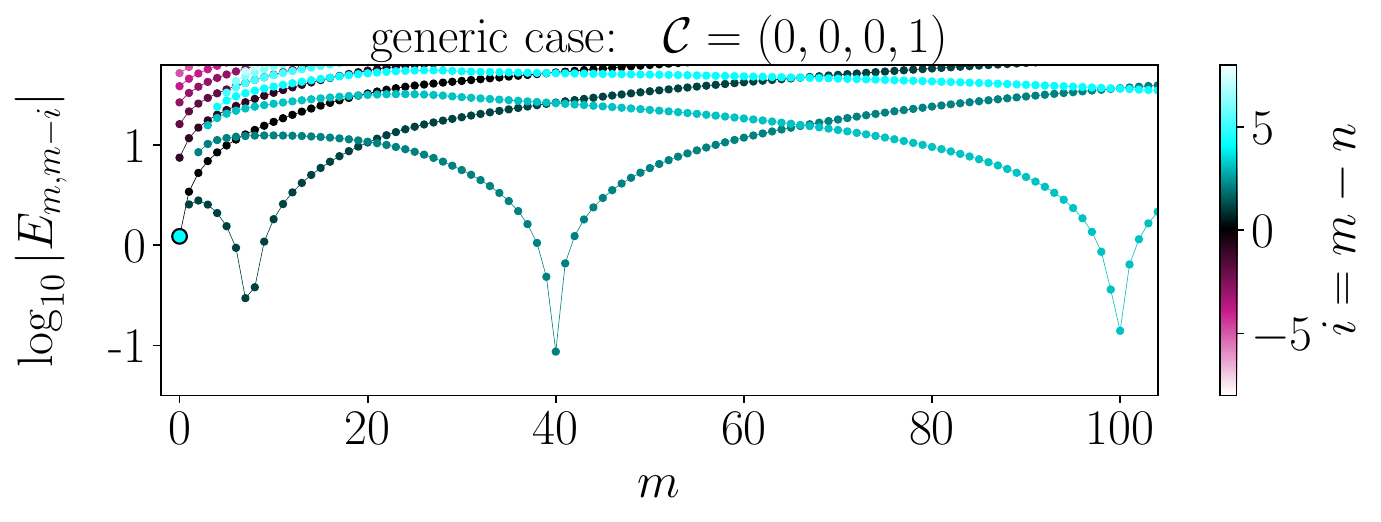}
    \\[-0.4em]
    \includegraphics[width=\linewidth]{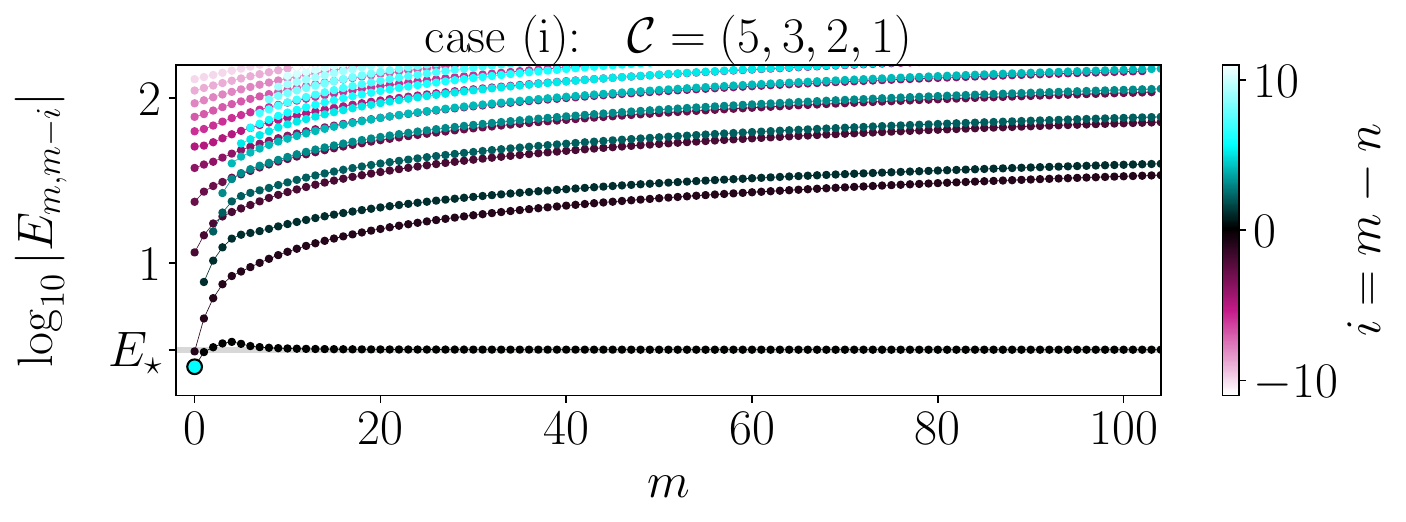}
    \\[-0.4em]
    \includegraphics[width=0.97\linewidth]{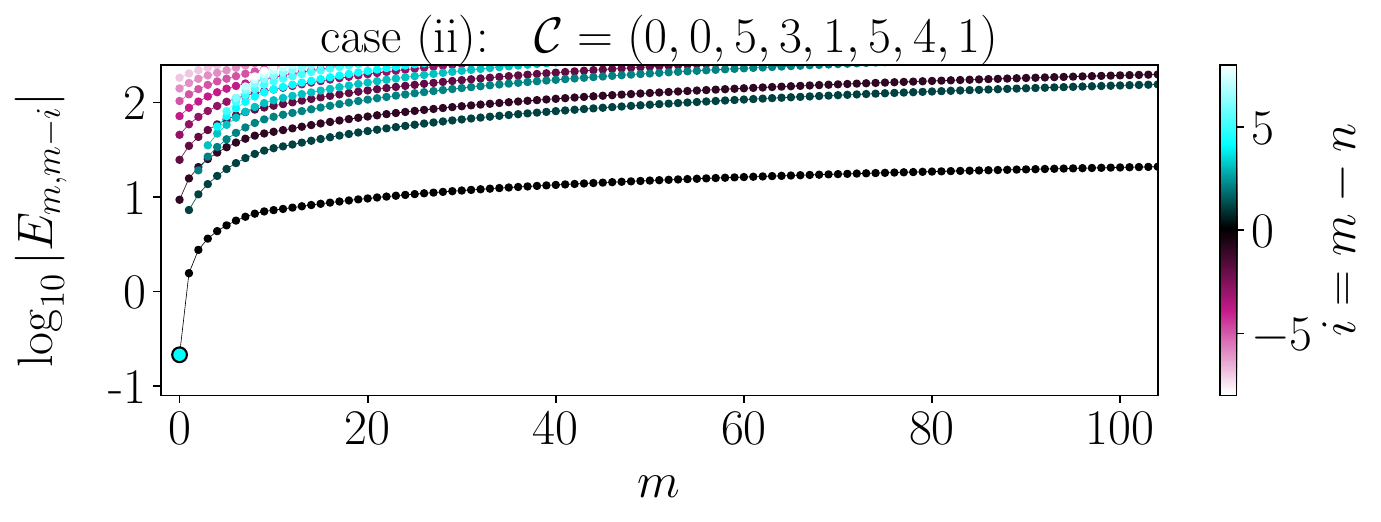}
    \\[-1.2em]
    \caption{
    Energy levels $|E_{m,m-i}|$ along diagonal sequences with offset $i{=}m{-}n$ for the polynomial subclass in~\cref{eq:polynomial-subclass-fg} with even boundary conditions ($\hbar\equiv1$). The upper panel exemplifies the generic case; the middle panel exemplifies case (i) with a single finite accumulation point at $E_\star$; the lower panel exemplifies case (ii) without any finite accumulation point. 
    The larger cyan dot marks $E_{00}{\neq}E_\star$.
    }
    \label{fig:accumulation}
\end{figure}

To illustrate the energy spectrum, we numerically compute energy levels for the polynomial subclass (see~\cref{eq:polynomial-subclass-fg}).
We discretise the kinetic term by symmetric second-order finite differencing and solve the resulting tridiagonal matrix eigenvalue problems on a sufficiently wide uniform grid of energy values. Interpolating these, we obtain the energy levels by a standard root-finding algorithm.
Examples are shown in~\cref{fig:generic-example,fig:accumulation}.
\\[-0.7em]

\noindent{\textbf{Asymptotic control of accumulation points.}}
Accumulation at $\bar E$ would require that for every $\epsilon>0$ the interval $(\bar E-\epsilon,\bar E+\epsilon)$ contains infinitely many energy levels $E_{mn}$. We will now show that standard one-dimensional Wentzel--Kramers--Brillouin (WKB) analysis suffices to identify coupling relations which ensure either (i) only a single accumulation point or (ii) no accumulation at all.

We determine the asymptotic eigenvalues $\lambda_m^{(u)}(E)$ as solutions to the Bohr--Sommerfeld quantisation condition
$S_u(\lambda,E)\,{=}\,\pi\hbar\,(m\,{+}\,\delta_u)\,{+}\,\mathcal{O}(1/m)$,
where $S_u(\lambda,E)\,{=}\,\int_{x_-}^{x_+}\sqrt{2(\lambda-V_u(x))}\,dx$ denotes the WKB action, evaluated between the classical turning points~$x_\pm$.
For the half-line problem, $x_-\,{=}\,0$ is fixed at the boundary and $\delta_u$ denotes a $\lambda$-independent phase, fixed by the inner boundary condition (Neumann $\delta_u\,{=}\,\tfrac{1}{4}$; Dirichlet $\delta_u\,{=}\,\tfrac{3}{4}$). For control of the remainder $\mathcal{O}(1/m)$, see e.g.~\cite{titchmarsh1962eigenfunction}.
An equivalent WKB analysis applies for $\lambda_n^{(v)}(E)$. 

We now define the continuous WKB action difference
\begin{align}
    \Delta S(\lambda,E)
    \equiv
    S_u(\lambda,E)-S_v(\lambda,E)
    \;,
\end{align}
for which we will determine the large-$\lambda$ behaviour below.
Once we subsequently impose the matching condition (see~\cref{eq:energy-quantisation-condition}), the two WKB quantisation conditions imply for $\Delta S$ (evaluated on these discrete solutions)
\begin{align}
    \Delta S(\lambda{=}\lambda_{mn},E{=}E_{mn})
    {=}
    \pi\hbar\left((m{-}n){-}\delta\right)\!
    {+}\mathcal{O}\!\left(\tfrac{1}{m},\tfrac{1}{n}\right)
    ,
    \label{eq:mismatch-condition}
\end{align}
with $\delta{=}\delta_u{-}\delta_v$.
We can obtain a convenient exact representation of $\Delta S(\lambda,E)$ by an interpolating potential
\begin{align}
    V_s(x)\equiv (1-s)V_v(x)+sV_u(x)
    \;,
\end{align}
with $s\,{\in}\,{[0,1]}$. The fundamental theorem of calculus gives
\begin{align}
    \Delta S
    &=
    -\frac{1}{\sqrt2}
    \int_0^1 ds
    \int_0^{x_+(s)}
    \frac{V_u(x)-V_v(x)}{\sqrt{\lambda-V_s(x)}}\,dx
    \;,
    \label{eq:letter-exact-DeltaS}
\end{align}
where $x_+(s)$ denotes the exact turning point, implicitly defined by $V_s(x_+(s)){\equiv}\lambda$ and the boundary term can be shown to vanish upon intermediate regularisation, i.e., $V_s(x_+(s)){\equiv}\lambda{-}\varepsilon$ with $\varepsilon{\to}0$.
Using the binomial representation of $\sinh^{2i}x$ and $\cosh^{2i}x$, the numerator becomes
\begin{align}
    \Delta V
    \equiv
    V_u{-}V_v
    =
    \hspace*{-1.5em}
    \sum_{\text{odd}\;1\leqslant p\leqslant N}
    \hspace*{-1.5em}
    A_p(\mathcal C)\,e^{2px}
    {-}
    \frac{E}{2}e^{2x}
    {+}
    \mathcal O(e^{-2x})
    \;,
    \label{eq:letter-M}
\end{align}
where all even orders cancel while the odd orders read
\begin{align}
    A_p(\mathcal C)
    =
    -2
    \sum_{k=p}^{N}
    (-1)^k\,
    \mathcal C_k\,
    \binom{2k}{k-p}\,
    \frac{1}{2^{2k}}
    \;.
    \label{eq:letter-Ap}
\end{align}
Since $A_p$ depends only on the $\mathcal C_{k\geqslant p}$, odd orders can be cancelled recursively by imposing $A_{N-1}\equiv0,\,A_{N-3}\equiv0,\,\dots$. This will be key to exclude accumulation and we henceforth denote with $p_*$ the largest remaining order.

The leading contributions to the denominator arise from $V_0(x){=}A_N e^{2Nx}$ (with $A_N{=}\tfrac{\mathcal C_N}{2^{2N}}$), for which the classical turning point is $x_+^{(0)}{=}\tfrac{1}{2N}\log(\lambda/A_N)$. Control of the subleading contributions requires to split the integration region into a bulk piece and a turning-point layer of width $w_\lambda{\sim}|x_+^{(0)}{-}x_+(s)|$. 
Overall, we find
\begin{align}
    \Delta S(\lambda,E)
    &=
    \hspace*{-1.0em}
    \sum_{\text{odd}\;1\leqslant p\leqslant N}
    \hspace*{-1.0em}
    K_p\,A_p(\mathcal C)\,\lambda^{\frac{p}{N}-\frac{1}{2}}
    -
    K_1\,\tfrac{E}{2}\,\lambda^{\frac{1}{N}-\frac{1}{2}}
    \notag\\*[-1.0em]&\quad\quad\quad\quad\quad\quad
    +
    \mathcal{O}\!\left(
        \lambda^{\frac{p_*}{N}+\frac{q_*}{2N}-1}
    \right)
    \;,
    \label{eq:key-estimate}
\end{align}
where we have also written out a bound for the subleading order of the remaining terms in the expansion of $V_s(x){=}V_0(x){+}\mathcal{O}(e^{2q_*x})$ with $q_*{<}N$. A detailed calculation of these estimates (including closed-form expressions for the constants $K_p$) will be presented elsewhere. 

Crucially, if we recursively cancel the odd $A_{p<p_*}$ down to some $p_*{<}N/2$, then~\cref{eq:key-estimate} implies that $\Delta S(\lambda,E){\to}0$ as $\lambda{\to}\infty$ (for any bounded $E$). 
Now consider any sequence of actual levels $(m_k,n_k,E_k)$ with $E_k{=}E_{m_kn_k}$ and $\lambda_k{\equiv}\lambda_{m_kn_k}{\to}\infty$. Along any such sequence, matching~\cref{eq:mismatch-condition} with~\cref{eq:key-estimate} implies $(m_k{-}n_k){-}\delta{=}0$ for all sufficiently large $k$. Since $m_k{-}n_k{\in}\mathbb Z$, this is only possible for equal parity, where $\delta{=}0$, and then only along the diagonal $m_k{=}n_k$. Thus, any finite-energy asymptotic sequence must eventually lie on this ridge.
We further distinguish two cases for the energy scaling on that ridge.
\vspace*{-1.4em}
\begin{itemize}
\setlength\itemsep{0.2em}
    \item[(i)] \textbf{Ridge accumulation.}
    If $p_*{=}1$, then~\cref{eq:key-estimate} takes the form $\Delta S\,{\sim}\,(E_\star-E)\,\lambda^{-X}\,{+}\,\mathcal{O}(\lambda_k^{-Y})$ with $0\,{<}\,X\,{<}\,Y$. Therefore, along any such asymptotic ridge sequence, matching with~\cref{eq:mismatch-condition} implies $E_{k}\,{=}\,E_\star + \mathcal{O}(\lambda_k^{X-Y})$ and thus $E_{k}\,{\to}\, E_\star$.
    \item[(ii)] \textbf{No accumulation.}
    If $1{<}p_*{<}N/2$, then~\cref{eq:key-estimate} takes the form $\Delta S{\sim}A\lambda^{-X}{+}E\lambda^{-Y}{+}\mathcal{O}(\lambda^{-Z})$ with $0{<}X{<}Y{<}Z$. Thus, along any such asymptotic ridge sequence, matching with~\cref{eq:mismatch-condition} forces $A\lambda_k^{-X}{\sim}{-}E_{k}\lambda_k^{-Y}$, hence $|E_{k}|\,{\sim}\,|A\lambda_k^{Y-X}|$, i.e, growing $|E_{k}|$ along any sequence as $\lambda_k\,{\to}\,\infty$.
    Thus, no finite accumulation point can occur. 
\end{itemize}
\vspace*{-0.4em}
We emphasise that all other remainder terms, both $\mathcal{O}(\lambda^{p_*/N + q_*/(2N) - 1})$ in~\cref{eq:key-estimate} and $\mathcal{O}(\tfrac{1}{m},\tfrac{1}{n})$ in~\cref{eq:mismatch-condition} are subleading, hence do not alter the above.

Of course, there remains the generic case with $p_*{>}N/2$, for which $\Delta S(\lambda,E)$ does not decay with $\lambda$ and the present analysis is insufficient to exclude accumulation. Clarifying this generic case is left for future work. 

For $N\,{=}\,4$, the $p\,{=}\,3$ term can be removed by tuning $\mathcal{C}_3\,{=}\,2\,\mathcal{C}_4$. The leading contribution becomes proportional to $E_\star\,{-}\,E$ (case (i)) and the spectrum exhibits a single finite accumulation point at $E_\star\,{=}\,\mathcal{C}_1\,{-}\,\mathcal{C}_2\,{+}\,\mathcal{C}_4$.

The same holds for $N\,{=}\,6$, where the odd orders $p\,{=}\,5,3,1$ appear. Both $p\,{=}\,5$ (growing with $\lambda$) and $p\,{=}\,3$ ($\lambda$-independent) need to be removed to avoid the generic case. After such cancellation, we again find case (i).

For $N\,{\geqslant}\,8$, case (ii) becomes possible. For instance, at $N\,{=}\,8$, the two growing contributions $p\,{=}\,7$ and $p\,{=}\,5$ can be removed by $\mathcal{C}_7\,{=}\,4\,\mathcal{C}_8$ and $\mathcal{C}_5\,{=}\,3\,\mathcal{C}_6\,{-}\, 14\,\mathcal{C}_8$. After this cancellation, the $p{=}3$ term leads to $\Delta S(\lambda)\,{\sim}\,A_3\lambda^{-1/8}$ which still dominates the energy-dependent contribution. We can thus exclude any finite accumulation point.
\\[-0.7em]

\noindent{\textbf{Discussion.}}
We have employed methods from separability theory to nonperturbatively quantise an integrable point-particle model with opposite-sign kinetic terms (ghosts) and genuine coupled interactions. Echoing previous results on classical stability~\cite{Deffayet:2021nnt,Deffayet:2023wdg}, we find that dominant self-interactions lead to bounded-below potentials in the two separated eigenvalue equations (see~\cref{eq:separated}), hence to discrete and bounded-below separated spectra $\lambda_m(E)$ and $\lambda_n(E)$. 
Energy quantisation follows from matching $\lambda_m(E)\,{\equiv}\,\lambda_n(E)$. The resulting energy spectrum $E_{mn}$ is unbounded above and below but not necessarily dense. In particular, we have constructed examples for which $E_{mn}$ either (i) admits exactly one accumulation point or (ii) no accumulation at all.
For the generic case, our analysis neither proves denseness nor excludes accumulation. 
Resolving this is left for future work.

The energy $E_{00}$ corresponds to the least excited separated state and can be considered as the shared ground-state energy. Details on the vacuum state and on unitary time evolution are presented separately~\cite{Deffayet:toAppear}. As we have numerically confirmed (see~\cref{fig:accumulation}) $E_{00}$ is generally distinct from any potential accumulation at $E_\star$. 

We highlight that we canonically quantise on the standard Hilbert space. Our quantisation is thus distinct from non-Hermitian or $\mathcal{PT}$-symmetric quantisation, see~e.g.~\cite{Bender:2007wu,Fring:2025zha}. Further, while St\"ackel systems generally admit bi-Hamiltonian constructions on an extended phase space with fiducial variables~\cite{gelfand1993local} (see~\cite{Kaparulin:2014vpa} and the review~\cite{ErrastiDiez:2024hfq} for application to higher-derivative systems), there is no reason to expect that the systems at hand are bi-Hamiltonian on the original phase space~\cite{Magri:1977gn}, see~\cite{fernandes1994completely} for the respective geometric criterion. It remains an open problem to extend our results to non-integrable models.

We also highlight concurrent developments on the stability of classical field theory~\cite{Deffayet:2025lnj,Held:2025fii}. 
In particular, for polynomial non-derivative interactions of sufficiently high degree, \emph{all} localised field configurations with sufficiently small initial data scatter and decay back to the vacuum~\cite{Held:2025fii}, see also~\cite{Figueras:2025gal} for a related conjecture. Here, dynamical decoupling occurs due to the dominance of geometric/dissipative decay of the free wave/Klein-Gordon equation. Moreover, radial scattering solutions in~\cite{Held:2025fii} suggest that the combination of this dominant decay at small data and an effective dynamical decoupling due to self-interactions at large data can even lead to global stability for all localised initial data of arbitrary amplitude.

Overall, the present work (see also~\cite{Deffayet:toAppear, Ewasiuk:toAppear}) thus fits into a tentative but coherent new picture: Ghost instabilities are not an inevitable \emph{kinematic} consequence of energy being unbounded above and below but rather depend also on whether or not the \emph{dynamics} allow for an uncontrolled energy transfer.
\\[-0.7em]

\paragraph*{Acknowledgements.}
We thank Christopher Ewasiuk and Stefano Profumo for sharing their results~\cite{Ewasiuk:toAppear}  prior to publication.
The work of S.~M. was supported in part by Japan Society for the Promotion of Science (JSPS) Grants-in-Aid for Scientific Research No.~24K07017 and the World Premier International Research Center Initiative (WPI), MEXT, Japan. 
The work of A.~V. and of A.~F.~J. was supported by European Structural and Investment Funds and the Czech Ministry of Education, Youth and Sports (Project FORTE CZ.02.01.01/00/22 008/0004632).

\bibliography{References}

\end{document}